\documentclass[conference]{IEEEtran}
\IEEEoverridecommandlockouts

\usepackage{cite}
\usepackage{amsmath,amssymb,amsfonts}
\usepackage{algorithmic}
\usepackage{graphicx}
\usepackage{textcomp}
\usepackage{xcolor}
\def\BibTeX{{\rm B\kern-.05em{\sc i\kern-.025em b}\kern-.08em
    T\kern-.1667em\lower.7ex\hbox{E}\kern-.125emX}}
\begin{document}

\title{CLIP-RL: Surgical Scene Segmentation Using Contrastive Language-Vision Pretraining \& Reinforcement Learning}

\author{
\IEEEauthorblockN{1\textsuperscript{st} Fatmaelzahraa Ali Ahmed}
\IEEEauthorblockA{\textit{Department of Surgery} \\
\textit{Hamad Medical Corporation}\\
Doha, Qatar \\
https://orcid.org/0000-0003-4534-2189}
\and
\IEEEauthorblockN{2\textsuperscript{nd} Muhammad Arsalan}
\IEEEauthorblockA{\textit{Department of Computer Science and Engineering} \\
\textit{Qatar University}\\
Doha, Qatar \\
https://orcid.org/0000-0003-1868-5207}
\and
\IEEEauthorblockN{3\textsuperscript{rd} Abdulaziz Al-Ali}
\IEEEauthorblockA{\textit{Department Computer Science and Engineering} \\
\textit{Qatar University}\\
Doha, Qatar \\
a.alali@qu.edu.qa}
\and
\IEEEauthorblockN{4\textsuperscript{th} Khalid Al-Jalham}
\IEEEauthorblockA{\textit{Department of Surgery} \\
\textit{Hamad Medical Corporation}\\
Doha, Qatar \\
kaljalham@hamad.qa}
\and
\IEEEauthorblockN{5\textsuperscript{th} Shidin Balakrishnan}
\IEEEauthorblockA{\textit{Department of Surgery} \\
\textit{Hamad Medical Corporation}\\
Doha, Qatar \\
https://orcid.org/0000-0001-6361-4980}
}

\maketitle

\begin{abstract}
Understanding surgical scenes can provide better healthcare quality for patients, especially with the vast amount of video data that is generated during MIS. Processing these videos generates valuable assets for training sophisticated models. In this paper, we introduce CLIP-RL, a novel contrastive language-image pre-training model tailored for semantic segmentation for surgical scenes. CLIP-RL presents a new segmentation approach which involves reinforcement learning and curriculum learning, enabling continuous refinement of the segmentation masks during the full training pipeline. Our model has shown robust performance in different optical settings, such as occlusions, texture variations, and dynamic lighting, presenting significant challenges. CLIP model serves as a powerful feature extractor, capturing rich semantic context that enhances the distinction between instruments and tissues. The RL module plays a pivotal role in dynamically refining predictions through iterative action-space adjustments. We evaluated CLIP-RL on the EndoVis 2018 and EndoVis 2017 datasets. CLIP-RL achieved a mean IoU of 81\%, outperforming state-of-the-art models, and a mean IoU of 74.12\% on EndoVis 2017. This superior performance was achieved due to the combination of contrastive learning with reinforcement  learning and curriculum learning.
\end{abstract}


\section{Introduction}
Minimally invasive surgeries (MIS) have replaced open surgeries in several specialties \cite{b16}, due to their substantial benefits, such as reduced blood loss \cite{b17}, diminished postoperative pain \cite{b18}, and faster recovery time \cite{b10}, \cite{b19}. MIS generates a vast amount of visual data via high-resolution surgical cameras \cite{b20}, offering unprecedented opportunities for enhancing surgical outcomes \cite{b54} and training through computer-assisted analysis \cite{b11}, \cite{b53}. However, the huge diversity of the scenes and the large size of the videos \cite{b22}, which may extend to hours at a time, make it labor-intensive \cite{b23} and time-consuming for data annotation \cite{b1}.

Recent advancements in segmentation methods indicate that traditional convolutional approaches have encountered a performance plateau. Most foundational models do not exceed a mean intersection over union (mIoU) of 75\% , such as S3Net \cite{b2}, MATIS Frame \cite{b3} and U-Net \cite{b4}. With the revolution of Vision-Language Models (VLMs), the Segment Anything Model (SAM) \cite{b5} has emerged as a state-of-the-art approach, excelling in mask generation. It has been fine-tuned for various domains, with many adaptations specifically tailored for the surgical field like Track Anything \cite{b7}, PerSAM \cite{b8}, and SurgicalSAM \cite{b9}. However, their reliance on prompts makes it impractical for analyzing lengthy surgical videos \cite{b12}.

To solve the issue of intensive segmentation labor, we propose a novel framework utilizing pre-trained VLMs for the segmentation of surgical instruments, thereby minimizing the requirement for manual annotations. Our model combines a ResNet-based CLIP model with a light-weight decoder and presents a reinforcement-learning-inspired adaptation mechanism.

\section{Background and Related Work}
Semantic image segmentation is widely used in medical imaging (e.g., tumor detection) \cite{b33} and robotics (e.g., surgical assistance) \cite{b35}. Traditional methods, which do not involve any learning process like thresholding and edge detection \cite{b36} struggle with complex textures and lighting variations. Deep learning-based approaches, particularly convolutional neural networks (CNNs) and vision transformers (ViTs), have significantly improved segmentation by capturing hierarchical and contextual image features.

\subsection{Convolutional Neural Networks  for Segmentation}
CNNs revolutionized segmentation by capturing spatial hierarchies. U-Net \cite{b4}, an encoder-decoder model, is widely used in medical imaging for precise segmentation \cite{b37, b38}. However, CNNs struggle with long-range dependencies and global context understanding.

\subsection{Vision Transformers for Segmentation}
ViTs \cite{b41} use self-attention to capture local and global features. ViTs are known for outperforming CNNs in tasks with large spatial contexts, such as whole-body MRI segmentation \cite{b42}. Models like Segmenter and SETR \cite{b43, b44} achieve state-of-the-art performance but require large datasets and high computational resources, limiting real-time applications.

\subsection{Hybrid CNN-Transformer Architectures}
Hybrid models combine CNNs for local feature extraction with transformers for global reasoning. TransUNet enhances medical segmentation (e.g., prostate MRI) \cite{b45, b46}. SegFormer replaces convolutional decoders with transformers, achieving robust results in diverse domains like aerial image segmentation \cite{b47}. 

\subsection{Vision-Language Models (VLMs) for Segmentation}
VLMs integrate textual descriptions with image data, improving generalization and interpretability. CLIP enables zero-shot segmentation, aiding rare tumor detection \cite{b48}. BLIP enhances self-supervised learning for medical annotation \cite{b49}. In robotic surgery, VLMs facilitate real-time anatomical highlighting, enhancing surgical decision-making \cite{b50}. These models push segmentation toward more interpretable AI-driven applications.

Our model adopts the frozen CLIP encoder as the backbone of an encoder–decoder segmentation network. CLIP’s contrastive learning paradigm clusters semantically similar images in feature space while pushing apart dissimilar ones, yielding robust, transferable representations \cite{b13,b9}.  Radford et al. (2021) demonstrate that CLIP achieves strong zero-shot transfer on over 30 classification datasets, often matching or exceeding task-specific models \cite{b48}.  By contrast, self-supervised ViTs such as DINOv2 \cite{b53} reach 85.8 \% top-1 accuracy on ImageNet in a linear-probe evaluation but do not leverage any language supervision \cite{b53}.  Importantly, in an industry benchmark spanning 117 classification datasets, CLIP-RN50×4-highres was the top performer in 80 \% of tasks versus 20 \% for DINOv2 \cite{b54}.  Thus, CLIP offers a truly prompt-free feature extractor—akin to pure-vision methods—while delivering superior transfer performance.

\begin{figure*}
\centerline{\includegraphics[width=1.2\columnwidth]{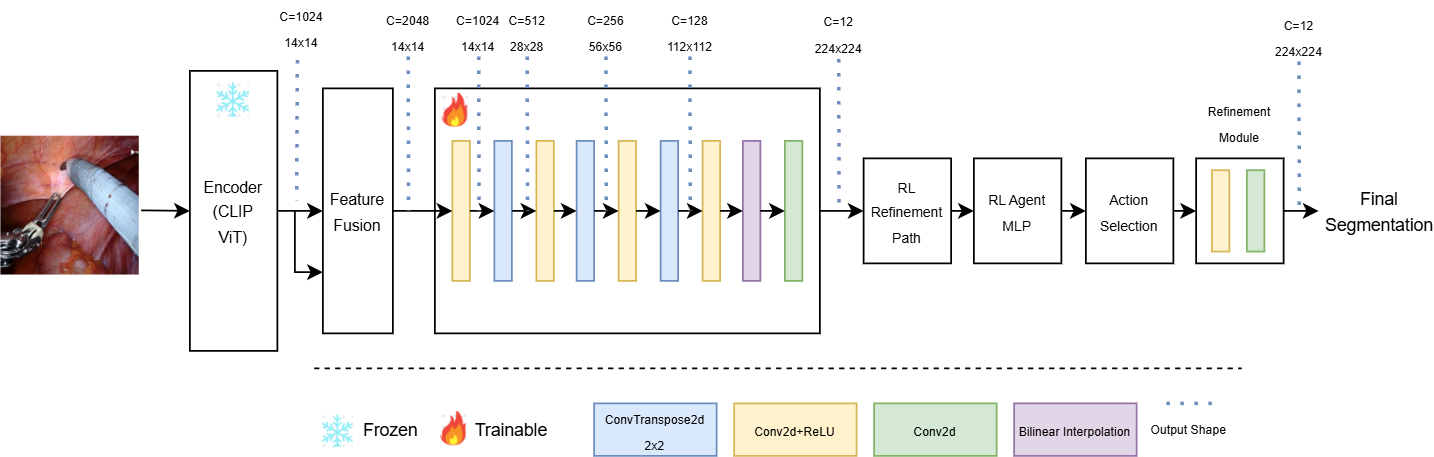}}
\caption{CLIP-RL Model Architecture}
\label{CLIP-RL Model Architecture}
\end{figure*}

\section{Methods}
\subsection{Methods Overview}

Our framework, \textbf{CLIP-RL}, integrates two key components, as shown in figure \ref{CLIP-RL Model Architecture}: a multi-resolution encoder-decoder segmentation network and an RL-based module. CLIP model acts as the model's encoder, which captures the input features. The generated feature map is passed to a lightweight decoder to generate the segmentation logit. Lastly, the RL-based refinement module functions as an adaptive decision maker that modulates the initial segmentation output by applying a residual correction \cite{b26}. The training process is guided by a hybrid loss function that combines conventional segmentation losses (cross-entropy and Dice losses) with a policy gradient loss derived from reinforcement learning. A curriculum learning strategy \cite{b28} is employed to gradually transition the training emphasis from segmentation specialization to full model collaboration to ensure robust performance in complex surgical scenarios.

\subsection{Detailed Design}
\textbf{Multi-Resolution Encoder–Decoder} The segmentation network adopts a multi-resolution encoder–decoder architecture, which is optimized for pixel-level prediction. The encoder is based on a pre-trained CLIP vision transformer that processes an input image $X$ and extracts hierarchical feature representations. The feature map is then passed through a feature‐fused ($F_{\text{fused}}$) module before entering the decoder. After discarding the CLS token, the remaining patch tokens are rearranged into spatial feature maps.
Then, the decoder reconstructs the segmentation map from $F_{\text{fused}}$ through a series of upsampling stages. For instance, a Conv2DTranspose layer upsamples the $14 \times 14$ feature map to $28 \times 28$ using a kernel size of 2 and a stride of 2, followed by convolutional refinement. This upsampling procedure is repeated until the spatial resolution of the output matches that of the input image, which is $244 \times $244 as shown in figure \ref{CLIP-RL Model Architecture}. The final segmentation output $s_L$ is computed as
\begin{equation}
s_L = Softmax(Conv(U_{P_{L-1}} \circ \cdots \circ U_{P_1}(F_{\text{fused}}))), \label{eq:segmentation}
\end{equation}
where $U_{P_l}$ denotes the Conv2DTranspose operation at level $l$. This formulation (see Equation  \eqref{eq:segmentation}) is especially well-suited for surgical segmentation, as it facilitates the recovery of fine anatomical details while leveraging deep semantic cues.

\textbf{RL-Based Refinement Module} Subsequent to the decoder, an RL-based refinement module is introduced to further enhance the segmentation predictions. A light-weight RL agent, implemented as a fully connected network, receives a globally pooled feature vector extracted from the final encoder output and generates a probability distribution over a discrete set of actions. Each action corresponds to a residual scaling factor; for example, the action space may consist of the set $\{-0.1, 0.0, +0.1\}$. Let $\alpha$ denote the scaling factor sampled from this distribution. Concurrently, an auxiliary refinement module computes a residual correction $r$ from the initial segmentation logits $s_L$. The final refined segmentation output $O$ is then formulated as:
\begin{equation}
O = s_L + \alpha \cdot r, \label{eq:refinement}
\end{equation}
 Equation \eqref{eq:refinement} encapsulates the adaptive adjustment mechanism. In the context of surgical segmentation, this refinement step is critical because even minor modifications in segmentation boundaries can have significant clinical implications, particularly in distinguishing between surgical instruments and adjacent tissues.

\textbf{Curriculum Learning for Hybrid Loss} To mitigate the potential instability of integrating the RL component, a learning strategy is employed in the curriculumlearning. The total training loss is defined as a weighted combination of the segmentation loss $L_{\text{seg}}$ and the RL loss $L_{\text{RL}}$. The segmentation loss typically comprises cross-entropy \cite{b29} and Dice losses \cite{b30}, \cite{b31}, ensuring accurate pixel-wise classification and overlap between predicted and ground-truth masks. The RL loss is derived from a policy gradient approach, computed as the negative log-probability of the selected action weighted by the advantage (i.e., the reward minus a running baseline). The total loss $L_{\text{total}}$ is given by
\begin{equation}
L_{\text{total}} = f_{\text{epoch}} \cdot L_{\text{seg}} + (1 - f_{\text{epoch}}) \cdot L_{\text{RL}}, \label{eq:total_loss}
\end{equation}
where the dynamic weighting factor $f_{\text{epoch}}$ is defined as
\begin{equation}
f_{\text{epoch}} = \left(1 - \frac{\text{epoch}_{\text{current}}}{\text{epoch}_{\text{total}}}\right)^2. \label{eq:curriculum}
\end{equation}

In the initial training epochs, $f_{\text{epoch}}$ is close to 1, thereby emphasizing the segmentation loss. As training progresses, $f_{\text{epoch}}$ decreases, which gradually shifts the focus toward the RL loss, promoting full model collaboration. Equations \eqref{eq:total_loss} and \eqref{eq:curriculum} formalize this strategy, ensuring that the model initially learns robust segmentation and subsequently refines its predictions through reinforcement learning—a progression that is particularly advantageous in high-stakes surgical segmentation scenarios.

\subsection{Implementation Details}

The framework is implemented in PyTorch to utilize the pre-trained CLIP vision transformer model. Input images are resized to $224 \times 224$ and converted to tensors. The CLIP encoder is maintained in a frozen state to preserve its robust semantic representations. After discarding the CLS token, the encoder extracts multi-resolution feature maps that are reshaped into spatial layouts, serving as the input for the decoder.

The decoder is designed as a multi-stage upsampling network. It employs a series of Conv2DTranspose layers (each configured with a kernel size of 2 and stride of 2) to progressively increase the spatial resolution (e.g., from $16 \times 16$ to $32 \times 32$, then to $64 \times 64$, and finally to $128 \times 128$). Each upsampling stage is followed by convolutional refinement blocks that integrate skip connections from the encoder, thus restoring fine-grained details. A final $1 \times 1$ convolution maps the refined feature maps to the desired number of segmentation classes, and a softmax activation yields voxel-wise probability distributions.

The entire model is trained using the Adam optimizer \cite{b32} with a learning rate of $1 \times 10^{-4}$. Moreover, model checkpoints are saved based on the highest observed mean Intersection over Union (mIoU) on a validation set, thereby preserving the best-performing model for clinical applications in surgical segmentation.

\section{Experiments \& Results}

We evaluated our proposed framework on two publicly available robot-assisted surgery datasets: EndoVis 2017 \cite{b14} and EndoVis 2018 \cite{b15}. The EndoVis 2017 dataset provides a well-established benchmark focused on tool segmentation in minimally invasive surgical procedures. EndoVis 2018 dataset offers a more comprehensive challenge by including annotations for surgical instruments and anatomical structures. In our study, we used EndoVis 2018 to perform a holistic scene segmentation. This two-task setting is critical as we need to accurately delineate the fine lines between surgical instruments and nearby tissues. We followed the training and test split of the data that is provided by the MICCAI community. For the validation, it was 20\% of the dedicated training dataset, which was splitted randomly. The model was trained on V100 on Paperspace by Gradient.

\subsection{Results}

We report our quantitative results using standard segmentation metrics, including the overall models' mean intersection over union (mIoU), per-class mIoU and Dice coefficient. Our Model outperformed SOTA models, even the ones that use prompts. 

\textbf{EndoVis 2017} dataset was used to evaluate our models' performance. It is a  robotic instrument segmentation dataset, which was part of a previous challenge. Our model was compared against state-of-the-art (SOTA) models: TransUNet \cite{b45}, SurgicalSAM \cite{b9}, PerSAM, and ISINet\cite{b52}.  Table \ref{tab:Endovis2017_results} presents the overall mIoU and perclass mIoU. CLIP-RL consistently outperformed baseline models, achieving an overall mIoU of 74.12\%. The model demonstrated exceptional performance across multiple instrument classes, achieving the highest mIoU in 5 out of 7 categories: PF (67.9), LND (78.43), GR (60.78), MCS (89.25), and UP (67.39). While S3Net recorded the highest mIoU for BF (75.0) and TrackAnything (5 Points) led in VS (83.68), our model secured the second-highest performance in these categories. The results highlight the advantage of our multi-resolution feature extraction, RL-based refinement, and curriculum learning strategy in improving segmentation precision.

\begin{figure}
\centerline{\includegraphics[width=0.7\columnwidth]{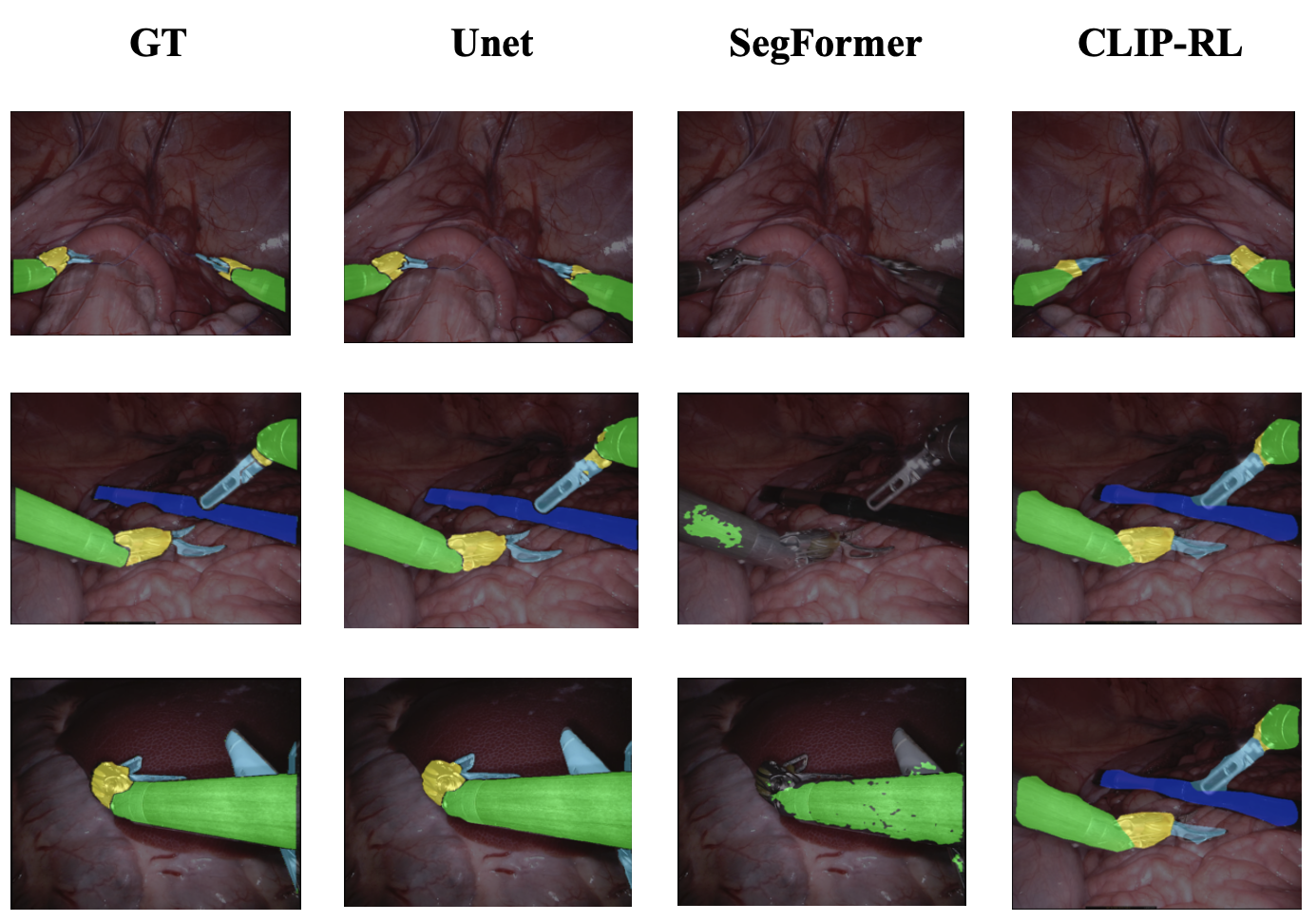}}
\caption{CLIP-RL Performance on EndoVis 2017 vs SOTA}
\label{Endovis 2017 Thubmnails}
\end{figure}

Figure \ref{Endovis 2017 Thubmnails} showcases qualitative segmentation results on EndoVis 2017. Compared to other models, CLIP-RL produces more precise instrument boundaries and exhibits greater robustness to variations in lighting, occlusions, and instrument overlap. Notably, the RL-based refinement module corrects segmentation errors from the initial prediction, reducing false positives and improving fine-grained delineation of surgical instruments. Our model outperformed SOTA models

\begin{table*}[ht]
    \centering
    \caption{Performance comparison of SOTA segmentation models vs CLIP-RL on EndoVis 2017. 
    \\
    BF: Bipolar Forceps, PF: Prograsp Forceps, LND: Large Needle Driver, VS: Vessel Sealer, GR: Grasping Retractor, MCS: Monopolar Curved Scissors, UP: Ultrasound Probe
    \\
    The Models' Results are Taken from SurgicalSAM\cite{b9}, The "*" Denotes That Model's Results are Reported by Us  }
    \label{tab:Endovis2017_results}
    
    \resizebox{0.7\textwidth}{!}{%
    \begin{tabular}{lcccccccc}
        \hline
        \textbf{Method} & \textbf{mIoU} & \textbf{BF} & \textbf{PF} & \textbf{LND} & \textbf{VS} & \textbf{GR} & \textbf{MCS} & \textbf{UP} \\
        \hline
        \textbf{SAM-based Models} & & & & & & & & \\
        Mask2Former + SAM & 66.21 & 66.84 & 55.36 & 83.29 & 73.52 & 26.24 & 36.26 & 45.34 \\
        TrackAnything (1 Point) & 52.46 & 47.59 & 28.71 & 43.27 & 82.75 & 63.10 & 66.46 & 55.54 \\
        TrackAnything (5 Points) & 64.50 & 55.42 & 44.46 & 62.43 & \textbf{83.68} & 62.59 & 67.03 & \textbf{65.17} \\
        PerSAM & 42.47 & 53.99 & 25.89 & 50.17 & 52.87 & 24.24 & 47.33 & 38.39 \\
        PerSAM (Fine-Tune) & 41.90 & 46.21 & 28.22 & 51.98 & 12.76 & 41.19 & 38.99 & 34.00 \\
        \textbf{SurgicalSAM} & 69.94 & 68.30 & 51.77 & 75.52 & 68.24 & 57.63 & 86.95 & 60.80 \\
        \hline
        \textbf{Specialist Models} & & & & & & & & \\
        TernausNet & 12.67 & 13.45 & 12.39 & 20.51 & 5.97 & 1.08 & 1.00 & 16.76 \\
        MF-TAPNet & 13.49 & 16.39 & 14.11 & 19.01 & 8.11 & 0.31 & 4.09 & 13.40 \\
        Dual-MF & - & 34.40 & 21.50 & 64.30 & 24.10 & 0.80 & 17.90 & 21.80 \\
        ISINet & 52.20 & 38.50 & 38.90 & 50.09 & 27.43 & 2.10 & 12.56 & 38.50 \\
        TraSeTr & - & 50.20 & 56.70 & 50.80 & 38.90 & 11.40 & 31.30 & 18.20 \\
        S3Net & 71.99 & \textbf{75.08} & 54.32 & 61.84 & 35.30 & 27.47 & 43.23 & 28.39 \\
        MATIS Frame & 62.74 & 66.18 & 50.99 & 52.23 & 32.84 & 15.71 & 19.27 & 23.90 \\
        \hline
        \textbf{CLIP-RL (Ours) *} & \textbf{74.12} & 70.12 & \textbf{67.90} & \textbf{78.43} & 75.42 & \textbf{60.78} & \textbf{89.25} & \textbf{67.39} \\
        \hline
    \end{tabular}%
    }
\end{table*}

\begin{figure}
\centerline{\includegraphics[width=0.8\columnwidth]{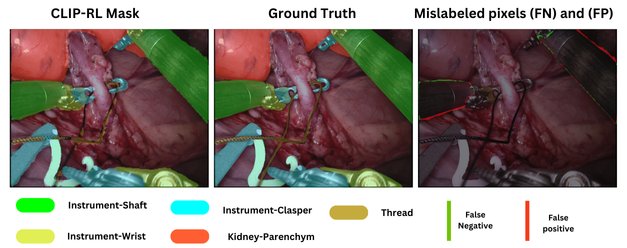}}
\caption{Qualitative segmentation results on EndoVis 2018 showing CLIP-RL masks, ground-truth, and false-positive/false-negative areas}
\label{Endovis 2018 Thubmnails}
\end{figure}

\textbf{EndoVis 2018} is a holistic surgical scene segmentation challenge, where both surgical tools and anatomical structures are segmented simultaneously. The results demonstrate the superior segmentation performance of CLIP-RL compared to state-of-the-art models. As shown in Table \ref{tab:performance}, CLIP-RL achieves the highest mean IoU (0.81) and Dice score (0.88), outperforming SegFormer (0.75 mIoU, 0.82 Dice), AdaptiveSAM (0.65 mIoU, 0.69 Dice), and nn-UNet (0.61 mIoU, 0.73 Dice).  Figure \ref{Endovis 2018 Thubmnails} shows our model's segmentation masks vs. ground truth. It can be seen that CLIP-RL segmented all instruments correctly except the thread, which was a challenging class for the model. 

 The per-class analysis in Table \ref{tab:per_class_results} further supports this, with CLIP-RL achieving the highest mIoU in 8 out of 11 classes, particularly excelling in instrument segmentation (IS: 0.93), clamps (0.85), and soft tissue structures like the small intestine (0.97). Additionally, it shows strong performance in IC, IW, SN, SI, and UP classes, reinforcing its generalizability. However, SegFormer outperforms CLIP-RL in CK (0.63 vs. 0.67) and Thread (0.58 vs. 0.66), suggesting areas for further optimization.  These results show that the overall performance of CLIP-RL surpasses U-Net, TransUNet, and SegFormer across most classes as well as the overall performance. These results highlight CLIP-RL as a SOTA solution for surgical video segmentation, demonstrating precise recognition of both instruments and anatomical structures.

\begin{table*}[ht]
    \centering
      \caption{Mean IoU and Dice values of CLIP-RL against SOTA models}
    \label{tab:performance}
    \renewcommand{\arraystretch}{1.2} 
    \fontsize{9}{9}\selectfont 
    \begin{tabular}{lcccccccc}
        \hline
        \textbf{} & \textbf{U-Net* \cite{b4} } & \textbf{nn-UNet} & \textbf{MedT} & \textbf{TransUNet} & \textbf{AdaptiveSAM} & \textbf{SAM \cite{b5}} & \textbf{SegFormer*} & \textbf{CLIP-RL (Ours)*} \\
        \hline
        \textbf{mIoU} & 0.58 & 0.61 & 0.64 & 0.52 & 0.65 & 0.62 & 0.75 & \textbf{0.81} \\
        \textbf{Dice} & 0.70 & 0.73 & 0.68 & 0.48 & 0.69 & 0.68 & 0.82 & \textbf{0.88} \\
        \hline
    \end{tabular}
\end{table*}

\begin{table}[ht]
\centering
\caption{Per-Class Segmentation Performance on EndoVis 2018 (mIoU) \\
IS: Instrument-shaft, IC: Instrument-clasper, IW: Instrument-wrist, KP: kidney-Parenchyma, CK: Covered-Kidney, SN: Suturing-Needle, SI: Suction-Instrument, S-Intestine: Small-Intestine, UP: Ultrasound Probe }
\label{tab:per_class_results}
\begin{tabular}{lcccc}
\hline
\textbf{Class ID} & \textbf{U-Net}  & \textbf{TransUNet} & \textbf{SegFormer} & \textbf{CLIP-RL (ours)} \\
\hline
\textbf{IS} & 0.65 & 0.67 & 0.69 & \textbf{0.93} \\
\textbf{IC} & 0.63 & 0.66 & 0.68 & \textbf{0.85} \\
\textbf{IW}& 0.64 & 0.67 & 0.70 & \textbf{0.83} \\
\textbf{KP} & 0.62 & 0.65 & 0.68 & \textbf{0.75} \\
\textbf{CK} & 0.61 & 0.64 & \textbf{0.67 }& 0.63\\
\textbf{Thread}   & 0.60 & 0.63 & \textbf{0.66 }& 0.58 \\
\textbf{Clamps}& 0.59 & 0.62 & 0.65 & \textbf{0.85} \\
\textbf{SN} & 0.58 & 0.61 & 0.64 & \textbf{0.77} \\
\textbf{SI} & 0.57 & 0.60 & 0.63 &  \textbf{0.83}\\
\textbf{S-Intestine} & 0.56 & 0.59 & 0.62 & \textbf{ 0.97}\\
\textbf{UP} & 0.55 & 0.58 & 0.61 &\textbf{ 0.89} \\
\hline
\end{tabular}
\end{table}

\subsection{Ablation Study}
To evaluate the impact of different components in our proposed framework, we conducted an ablation study by incrementally adding key modules: Curriculum Learning and Reinforcement Learning (RL). The baseline model without these components serves as the reference. 

The results, presented in Table \ref{tab:ablation}, show that the addition of Curriculum Learning improves both mIoU and Dice scores by providing a progressive training strategy that enhances model convergence on complex regions. Further integrating RL enables the model to dynamically adapt its predictions, leading to additional performance gains.

\begin{table}[ht!]
\centering
\caption{Ablation study of the proposed model based on mIoU and Dice metrics.}
\label{tab:ablation}
\begin{tabular}{lcc}
\hline
Model Configuration              & mIoU (\%) & Dice (\%) \\
\hline
Baseline                         & 72.4      & 75.1      \\
Curriculum Learning            & 76.8      & 79.3      \\
Curriculum Learning + RL       & \textbf{81.0 } & \textbf{88.0}  \\
\hline
\end{tabular}
\end{table}

\section{Conclusion}

In this work, we introduced CLIP-RL, a novel framework for surgical image segmentation that integrates a multi-resolution encoder-decoder network with an RL-based refinement module. By leveraging a frozen CLIP vision transformer, our model effectively captures both fine-grained spatial details and high-level semantic information, ensuring accurate segmentation of anatomical structures and surgical instruments. The reinforcement learning (RL) module further refines segmentation predictions through an adaptive residual correction mechanism, allowing for dynamic adjustments that enhance segmentation precision. 

To ensure training stability and maximize performance, we employed a curriculum learning strategy that gradually transitions the learning emphasis from segmentation accuracy to full model collaboration. This approach effectively balances conventional segmentation losses (cross-entropy and Dice losses) with policy gradient optimization, allowing the RL agent to refine predictions in a structured manner.

Experimental results demonstrate that CLIP-RL outperforms existing state-of-the-art models in surgical segmentation tasks, achieving superior results in terms of both mIoU and Dice scores. The combination of vision-language pretraining, reinforcement learning, and curriculum learning makes CLIP-RL particularly well-suited for the challenges of surgical video analysis, where precision and adaptability are critical.

Future work will include extending our approach to multi-modal fusion and incorporating additional surgical cues, such as temporal video information and instrument kinematics, and improve segmentation accuracy in dynamic surgical environments.

\section*{Acknowledgment}

The authors would like to acknowledge the support of the Surgical Research Section at Hamad Medical Corporation for the conduct of this research

Research reported in this publication was supported by the Qatar Research Development and Innovation Council (QRDI) grant number ARG01-0522-230266. The content reported through this research is solely the responsibility of the authors and does not necessarily represent the official views of Qatar Research Development and Innovation Council

\end{document}